# Non-specific cellular uptake of surface-functionalized quantum dots

T A Kelf[1], V K A Sreenivasan[1], J Sun[1], E J Kim[2], E M Goldys[1] and A V Zvyagin[1]

[1] MQ Photonics Centre, Faculty of Science, Macquarie University, Sydney, Australia

[2] Department of Science Education-Chemical Education Major, Daegu University, Gyeonbuk, South Korea

E-mail: azvyagin@science.mq.edu.au

**Abstract.** We report a systematic empirical study of nanoparticle internalization into cells via non-specific pathways. The nanoparticles were comprised of commercial quantum dots (QDs) that were highly visible under a fluorescence confocal microscope. Surface-modified QDs with basic biologically-significant moieties, e.g. carboxyl, amino, streptavidin were used, in combination with the surface derivatization with polyethylene glycol (PEG) in a range of immortalized cell lines. Internalization rates were derived from image analysis and a detailed discussion about the effect of nanoparticle size, charge and surface groups is presented. We find that PEG-derivatization dramatically suppresses the non-specific uptake while PEG-free carboxyl and amine functional groups promote QD internalization. These uptake variations displayed a remarkable consistency across different cell types. The reported results are important for experiments concerned with cellular uptake of surface-functionalized nanomaterials, both when non-specific internalization is undesirable and also when it is intended for material to be internalized as efficiently as possible.

**Introduction**

To sustain the lifecycle of living cells material must be absorbed from the extracellular medium via certain mechanisms. These mechanisms fall into two categories: specific internalization, that requires the cell to actively recruit molecules into the cytoplasm, and non-specific internalization, which constitutes random processes in which the cell has no active control. Receptor-mediated endocytosis is an important example of the specific uptake mechanism that facilitates import of selected extracellular macromolecules and allows inter-cellular signalling. This process is regulated by plasma membrane receptors that are only activated by receptor-specific ligands [1-4]. As a result, only biomolecular complexes grafted with these ligands can gain entry into the cell. In the process of endocytosis the plasma membrane is engulfed inwards from specialized membrane micro-domains forming either clathrin- or caveolin-coated pits. In recent years this specific cellular uptake mechanism has been a subject of intense research driven by the motivation to better understand cellular molecular trafficking and potential applications in targeted drug delivery [5]. Non-specific cellular uptake refers to a process of extraneous material internalization; it can vary depending on the cell type and has poor material selectivity. For virtually all eukaryotic cells, non-specific uptake is dominated by pinocytosis [6, 7]. This process involves continual budding of vesicles from the cell surface, allowing fluid and extra-cellular material to enter the cell. After either specific or non-specific internalization the material trapped in the vesicles is sequestered into endosomes, subsequently it is transported to locations within the cell [4], moved back to the extra-cellular medium, or taken to the lysosomes for acidic degradation. Nanoparticles incubated with cells are eventually concentrated in lysosomes or in the perinuclear recycling compartment [8] as they are of no use to the cell. Often, due to the similarity of pinocytosed and receptor-mediated internalization patterns, the discrimination of the non-specific versus specific uptake processes is not straightforward. The literature reports addressing non-specific cellular uptake and the key factors determining its rate are dispersed over various research fields from gene transfection [9, 10] and drug delivery [11, 12] to nanoparticle synthesis [13, 14]. They provide evidence that the key factors determining internalization are size [15, 16, 17], charge [18, 19] and surface functional groups [12], where polyethylene glycol (PEG) is of particular importance [20-22].

Nanoparticles serving as biomolecular cargo vehicles hold considerable promise for targeted delivery, with their large surface areas hosting various functional moieties that can dock different biomolecules. Quantum dots (QDs) have been intensely investigated due to their exceptional luminescent properties: unprecedented efficiency (high action cross-section, which is a product of quantum efficiency and absorption cross-section [23]). Spectral emission tunability by size variations allowing spectral multiplexing and their convenient excitation band in the UV spectral range make QDs particularly attractive as fluorescent labels and they have already been widely used in a number of biological studies, as molecular reporters [24]. Their large two-photon action cross-sections compared to the best organic fluorophores add further to their appeal [23]. The QD technology is maturing rapidly with a range of QD products available commercially, including QDs with amino, carboxyl and streptavidin surface groups. The toxicity of QDs remains a hotly debated issue, which may limit medical applications [25-28], however their brightness and photostability make them ideal for *in-vitro* biomedical applications.

In this paper, we investigate the cellular internalization rates of a range of different commercially available QDs including streptavidin-coated QDs and carboxyl- and amino- surface functionalized QDs with and without PEG linker chains. Carboxy- and amino- functionalised QDs have been chosen because they are readily amenable to further conjugation with biomolecules following various chemical treatments. Streptavidin-coated QD are important because, by virtue of the exceptional affinity of streptavidin for biotin (dissociation constant ~ $10^{-15}$ M), streptavidin QDs can be easily attached to biomolecules of interest (ligands/proteins/antibodies), which have been coupled to biotin. Some of the QDs investigated here were also modified chemically to further understand the effect of chemical groups on cellular internalisation dynamics. This modification revealed significant modulation of the cellular uptake rates.

The aim of this paper is to provide an initial guide for the use of nanoparticles in cellular systems. While this work focuses specifically on QDs, the results and discussion can be applied to various other types of nanoparticles as the effects of size, surface charge and molecular coatings are generally more important than the chemical identity of nanoparticles. This general discussion aims to shed light in issues arising in early stages of the research on application of nanoparticles in cellular systems. Our experience suggests that without proper controls and time-consuming investigations the reasons for particle internalization can be hard to elucidate.

**Experimental Procedure**

*1. Quantum Dots*

All QDs used in this study were obtained from two major suppliers: Invitrogen (Oregon, USA), (referred to as $QD_i$) and eBiosciences (San Diego, USA), (referred to as $QD_e$). Both have the same basic nanostructure consisting of a cadmium selenide core passivated with a layer of zinc sulphide (CdSe/ZnS). For this study, QDs with a fluorescence emission maximum at 605 nm were chosen; the QD excitation band is broad, mostly in the ultraviolet spectral range, and has an additional excitation peak at 545 nm allowing convenient excitation in visible. All QDs are capped with an amphiphilic polymer coating by the manufacturer. This coating serves three purposes: to reduce cytotoxicity, improve colloidal stability, and enable the grafting of functional moieties. Hence, the overall size of these QD complexes ranged from 15 nm to 25 nm depending on the surface functional groups. This work used seven different types of QDs whose full list is shown in figure 1. They were, as follows: Carboxyl-QDs (referred to as $C-QD_i$), Amino-PEG QDs ($N-pQD_i$, where p refers to the presence of polyethylene-glycol, PEG) and Streptavidin-QDs ($SA-QD_i$), all purchased from Invitrogen. Carboxyl-PEG QDs ($C-pQD_e$) and Amino-PEG QDs ($N-pQD_e$), were purchased from eBiosciences. To further understand the effect of the surface terminal groups, $C-QD_i$ and $C-pQD_e$ were conjugated to ethylene-diamine to obtain amine-terminated quantum dots, referred to as $N-CC-QD_i$ and $N-CC-pQD_e$. The QDs can be divided into PEG-derivatized and PEG-free, these are drawn in the left and the right column in figure 1, respectively. It is also possible to classify the QDs with respect to the chemical terminal group (amine, carboxyl or streptavidin), some of which have a reacted variant.

For the coupling reaction 400 fmole (1 molar equivalent) of QDs were diluted in 400 µL (MES) buffer with 2 mM (EDTA, pH=6) in a vial. Freshly dissolved room temperature equilibrated 1-ethyl-3-(3-dimethylaminopropyl) carbodiimide (EDC) (10 mg/mL) and sulfo- N-Hydroxysuccinimide (sulfo-NHS) (20 mg/mL) solutions in deionized water were added to the QD vial for activation and the solution was incubated at room temperature with gentle shaking for 15 minutes. An excess of ethylene-diamine was added along with 50 μL of 5x borate buffer (pH = 8.4). After 1-hour room temperature incubation with gentle shaking free EDC, sulfo-NHS and ethylene-diamine were removed by using a desalting column (Prepacked Sephadex G-25 medium column, GE Healthcare) equilibrated with phosphate buffer saline PBS (pH = 7.2).

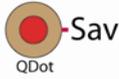

*Figure 1. List of commercially available QDs from Invitrogen and eBiosciences. Chemical structure of the surface groups is shown along with the short hand notation. Additional chemical reactions have been performed where noted.*

the concentration and quantum efficiency of the reacted QDs were measured using an absorption spectrometer and fluorimeter, by comparing with untreated QDs. We found that Invitrogen QDs have approximately 2.5 times greater quantum efficiency than that of the eBiosciences QDs. After the EDC reaction the Invitrogen QDs (N-CC-QD$_i$) and eBiosciences QDs (N-CC-pQD$_e$) suffered a 2.5- and 10- fold drop in quantum efficiency, respectively.

*2.    Cell Experiments*

Two rat tumour cell lines were obtained from ATCC (USA), these were GH4C1 (pituitary) and AR42J (pancreas) cells. A pituitary tumour cell line, AtT20 was a kind gift from Dr M Connor (The University of Sydney, Australia). Ar42J cells were cultured in F-12K medium (DKSH) with 20% fetal bovine serum. Both GH4C1 and AtT20 cells were cultured in F-10 Ham's (DKSH) with 2.5% fetal bovine serum and 15% horse serum. All cells were grown in culture flasks (BD Falcon) inside a humidified incubator at 37$^o$C with 5% $CO_2$.

The day before experiments cells were transferred to 8 chambered culture slides (BD Falcon) and incubated in medium, as above. The cells were then incubated with a 20-nM QD solution in 300-μL phosphate buffer saline (PBS, Invitrogen, pH = 7.2) with additional $CaCl_2$ (0.9 mM), $MgCl_2$ (0.5 mM), BSA (0.1%) and D-Glucose (20 mM) at 37$^o$C, 5% $CO_2$ for 60 min. After QD incubation, cells were washed twice with PBS then fixed using 3.7% (w/v) paraformaldehyde solution for 20 min followed by a final PBS wash.

*3.    Imaging and image processing*

A confocal laser-scanning microscope (Leica TCM SP2) with a 488-nm Ar-ion laser was used for acquisition of fluorescence images of the QDs. An external Nikon DS-Qi1Mc camera was attached to acquire the respective bright field, transmission images. The internalization was quantified by fluorescence intensity of the QDs inside the cells. For each experiment, approximately 20 cells were imaged using a spectral filter of the bandwidth 585 nm – 625 nm to accommodate for the QD emission and reject illumination light. All imaging parameters were kept constant for all experiments to allow direct comparisons. After background correction and intensity scaling to account for different QD quantum efficiencies, the total brightness inside each cell was calculated (software Igor Pro, Wavemeterics). These values were normalized to the cell size and averaged over all cells to provide a single number that corresponded to the level of internalization for each QD type into each cell line. These values were subsequently normalized to the internalization rate of N-CC-QD$_i$ into AtT20 cells to facilitate comparison between different cell lines and QD types.

**Results and Discussion**

*Results*

Firstly, we examined the internalization mechanism of QDs. Figure 2a shows representative images of 20 nM of C-QD$_i$ incubated with Ar42J cells for one hour at 4$^o$C and 37$^o$C. No internalization was observed at 4$^o$C, with the QDs localized on the cell membrane. The cellular machinery was switched on at 37$^o$C manifested by the QD uptake in the cellular cytoplasm at the expense of the reduced concentration of QDs localized on the membrane. This internalization mechanism was active, as opposed to, for example, the passive diffusion cellular uptake of small-size molecules [29]. This observation is consistent with receptor-mediated endocytosis [30] that is not effective at 4°C. The receptor-mediated endocytosis mechanism of the QD uptake was additionally confirmed by pre-treatment of the Ar42J cells with sucrose before the incubation of 20 nM of C-QD$_i$, with the result shown in figure 2b. In this case, no internalization was observed. Sucrose is known to disrupt the formation of clathrin vesicles [16, 31], hence this shows that C-QD$_i$ were predominantly internalized via endocytosis that involved clathrin-coated vesicles. Figure 2c shows the detected level of C-QD$_i$ in Ar42J cells as a function of incubation time. The internalization rates were calculated by analysing the number of QDs inside the cells (ignoring those localized on the membrane), and averaging over 20 cells for each time point. The internalization saturated at approximately 10 minutes, a period limited by the number and activity of the receptors recruited for the internalization process. The variation in the internalization level at longer incubation times is explained by statistical variations in the cell uptake. We verified that there was no observable variation in the internalization rate up to 90 minutes. The incubation time of 60 minutes was used in further experiments to ensure that the internalization of the nanoparticles was as complete as possible.

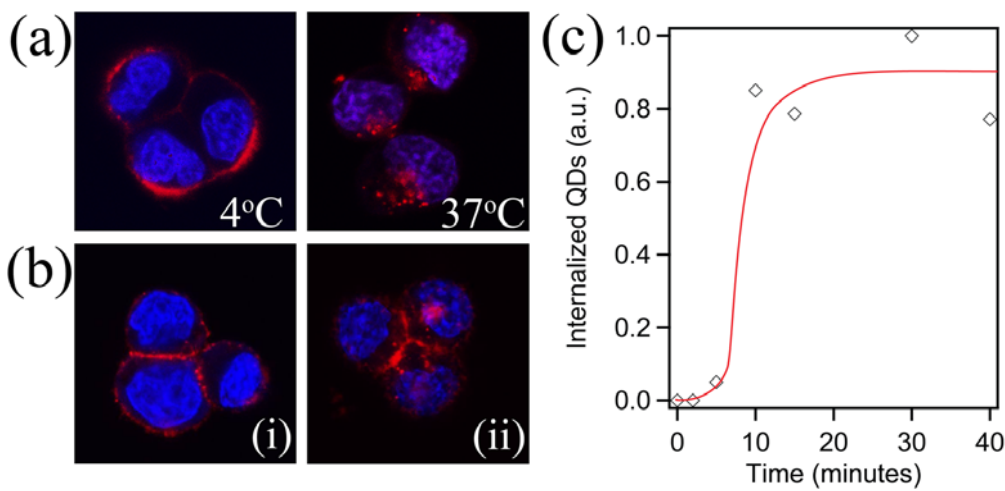

*Figure 2. (a) C-QD$_i$ incubated at 4°C and 37°C in Ar42J cells, QDs are pseudo-colour-coded in red, nuclei were stained with blue fluorescent dye Hoerst and shown in blue colour. Image size 37.5 μm × 37.5 μm. (b) C-QD$_i$ incubated (i) with sucrose and (ii) without sucrose in Ar42J cells. Image size and colours are the same as in (a). (c) plot of the C-QD$_i$ internalization into Ar42J cells versus incubation time.*

Figure 3 shows representative images of different QDs in the tested cell lines. Figure 4 shows their calculated internalization levels, with the values summarized in table 1. QD sizes and zeta potentials, measured using a dynamic light scattering (DLS) method (Zetasizer, Malvern, UK) are also presented in table 1. QD hydrodynamic diameters (which overestimate the geometric average diameter by roughly 10%) were measured to be approximately 20 nm with the exception of N-CC-QD$_i$ (100 nm, table 1). This larger measured size reflects the process of QD aggregation in the buffer solution. We have tested all QD solutions in water and PBS a number of times and have sometimes found aggregates present in the data. The use of PEG and the amphiphilic polymer coating act to reduce aggregation significantly over that of bare particles but never completely remove the effect. Hence, sizes of the nanoparticle complexes formed in buffer medium, rather than their primary sizes, should be taken into account in the context of cellular internalization. The effect of nanoparticle size will be discussed later; however we note that QD aggregates were under 200 nm in size, with smaller aggregate/primary particle often present in buffers. Such aggregation levels still permitted the efficient QD uptake into the cells, as shown in figure 3 for N-CC-$QD_i$ since the vesicles can accommodate up particles up to ~ 200 nm.

We now discuss the general trends that can be seen in the data. On average, Ar42J and GH4C1 cells had a similar level of internalization, which was, roughly, half that of AtT20 cells. QDs surface-derivatized with PEG showed comparatively low levels of cellular internalization. We note internalization of N-$pQD_i$ was virtually negligible, at the noise level of the detection. The level of the SA-$QD_i$ internalization was slightly greater than that of most of the PEG-derivatized QDs, except for AtT20 where low internalization rate is observed.

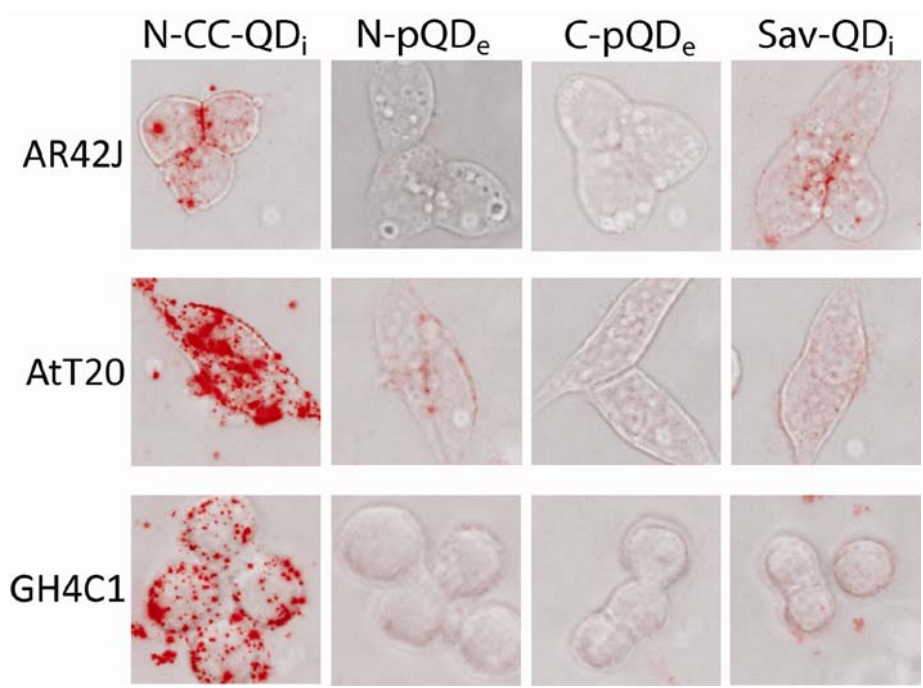

*Figure 3. Representative bright-field images of the tumour cells superimposed with the epifluorescence images of quantum dot fluorescence (red). The image acquisition parameters were constant for all experiments to facilitate direct comparison. Each image panel size 37.5 μm × 37.5 μm.*

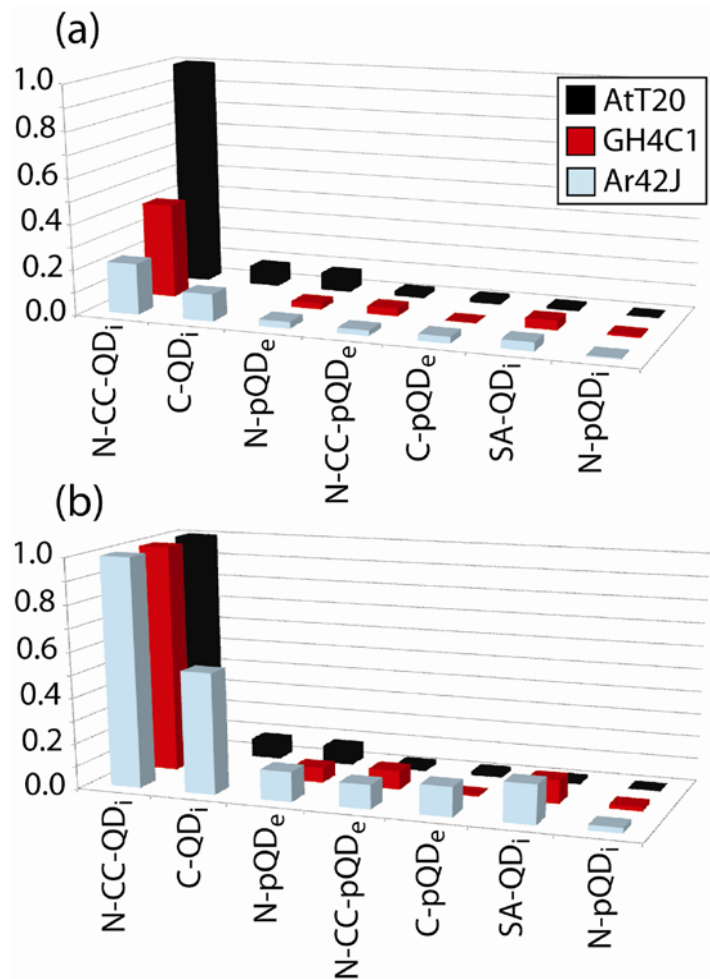

*Figure. 4. Bar plot of the QD uptake levels for the three tumour cell lines: (a) normalized to that of N-CC-QD$_i$ quantum dots in AtT20 cells; (b) normalized across each cell line data to that of N-CC-QD$_i$.*

*Table 1. Normalized QD uptake rates into the 3 tumour cell lines, QD size and zeta potential.*

| QD Type | Internalization | | | Size (nm) | Zeta Potential (mV) |
|---|---|---|---|---|---|
| | Ar42J | GH4C1 | AtT20 | | |
| N-CC-QD$_i$ | 0.225 | 0.419 | 1 | 100 | -32 |
| C-QD$_i$ | 0.12 | - | 0.082 | 28 | -31 |
| N-pQD$_e$ | 0.03 | 0.027 | 0.078 | 23 | -2 |
| N-CC-pQD$_e$ | 0.024 | 0.035 | 0.022 | 24 | -12 |
| C-pQD$_e$ | 0.029 | 0.002 | 0.018 | 22 | -12 |
| SA-QD$_i$ | 0.039 | 0.043 | 0.012 | 20 | -15 |
| N-pQD$_i$ | 0.005 | 0.008 | 0.002 | 21 | -2 |

*Discussion*

It is well known that a number of factors affect the rate of non-specific particle internalization into cells. These include charge [18, 19], size [15, 16, 17], and surface functional groups [12]. Additionally, hydrophobic nanomaterials which have an affinity to the lipid bilayer of the cell facilitate uptake [32]. Pristine amphiphilic-polymer-coated QDs and PEG-derivatized QDs have similar hydrophilic properties, and yet as shown, their cellular uptakes are dramatically different. To understand this difference we will discuss the roles of the receptor-specific mediated pathway, size effects, PEG-derivatization, and charge, in the internalization process.

Firstly, we discuss whether direct receptor mediated pathways plays a significant role in the internalization of the streptavidin-coated QDs in comparison with non-specific mechanisms. Streptavidin has a high binding affinity to biotin, which belongs to the class of vitamins. Since biotin is present in the culture medium and can be adsorbed onto the cell membrane this could enhance the internalization of Sav-$QD_i$ through streptavidin biotin binding [33]. Also, it has been reported that the RYD peptide chain in streptavidin is sufficiently similar to the sequence RGD that is used by many cell adhesion-related molecules, and, thus, unmodified streptavidin can be directly internalized [34]. Thirdly, there can be non-specific internalization through pinocytosis, a common pathway for a broad class of extraneous material internalisation, including QDs. We tested the first pathway by pre-binding biotin to the streptavidin-coated QDs, and found no discernable change in the level of internalization. This ruled out the possibility of internalization via the biotin-mediated pathway. The second pathway was more difficult to test, however the total observed internalization levels of the streptavidin-coated QDs were broadly comparable to the lowest levels of internalization of the other (PEG-containing) QDs, that were limited to the pinocytosis pathway. This suggests that pinocytosis was the dominant mechanism. The added complexity of the potential multiple internalization pathways makes streptavidin-coated QDs less attractive, as a reference nanoparticle unless preliminary work is conducted to ensure the receptor-mediated internalization is significantly greater compared to alternative processes.

With respect to the carboxyl group mediated internalization, it has been shown that carboxylic QDs were endocytosed via lipid rafts [35]. However, since the sucrose-induced inhibition of the C-$QD_i$ internalization strongly suggested the clathrin-vesicle pathway (figure 2b), the lipid raft internalization mechanism is ruled out as a dominant factor for these cell lines. At the same time, its active role in the reported internalization scenarios [35] indicates that the process can be more complex and makes it difficult to explain C-$QD_i$ internalization within a simple framework.

Size has been shown to have a strong influence over the rate of nanoparticle internalization. Whilst our primary QDs were of a similar size, it was noted that there can be aggregation in the culture medium, leading to distribution of actual sizes. The effect of size on internalisation has been addressed in several papers [16, 17, 37-39], and although consensus has not been reached, a general trend is emerging, as schematically shown in Figure 5. The size effects fall approximately into three classes: Clathrin, Caveolae, Macro pinocytosis.

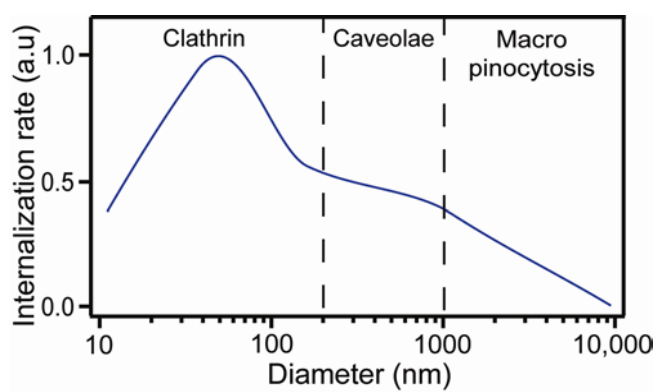

*Figure 5. Qualitative plot of the internalization level versus the particle size. This plot was compiled using published data [16, 17, 37-39] obtained for a variety of cell models and particle types.*

Clathrin-mediated endocytosis pathways are predominant for nanoparticles smaller than 200 nm, and it has been shown both experimentally [16] and theoretically [38] that 50 nm particles are most efficiently endocytosed. The non-specific internalization in this size range is hypothesized to occur via adsorption of the proteins from the culture medium onto the particles surface, followed by their internalization via receptor-mediated pathways. Thus the rate of internalization is dependent on the available receptors on the cell surface and the number of receptors a single nanoparticle triggers when it adheres to the membrane. It has been shown experimentally that for particles <50 nm, several closely spaced particles are required to trigger the formation of a clathrin-coated vesicle [17]. As a result, an internalization rate per particle is reduced, as manifested by a descending slope at sizes below 50 nm. A theoretical model based on the balance between free energy required to drive nanoparticles into the cell and diffusion kinetics of the recruitment of receptors to the binding site [38] helps explain this decreasing internalization level. For particle sizes in excess of 50 nm, more receptors are engaged by the particle internalization process followed by their translocation into the cytoplasm. To sustain the internalization, these receptors must be recycled back to the membrane, which takes time and limits the internalization rate. When nanoparticles reach sizes greater than 200 nm they can no longer be internalized through clathrin vesicles, and cells recruit alternative internalization pathways. It has been shown [17] that caveolae-mediated pathways were of the greatest importance for particles between 200 nm and 1000 nm. This process is slower than that of the clathrin internalization, hence the efficiency is lower. For particles greater than a micron in size (up to 5 μm in diameter), the effect of macro-pinocytosis becomes important. This requires ruffling of the cell membrane to form large troughs, which can subsequently bud in the form of large size vesicles into the cell. This is a genuinely non-specific process, and highly dependent of cell type and internalization conditions. The internalization rate drops dramatically above around one micron due to difficulty of forming these large vesicles.

Particle shape has also been shown to play a role in the internalization [37], which adds another layer of complexity. We note briefly that a cell perceives the particle in terms of its surface area in contact with the cell membrane, and thus this size governs internalization, as discussed above.

We now turn to the role of PEG in the non-specific nanoparticle uptake in cells. While precise reasons behind the significantly reduced internalization of PEG-derivatized nanoparticles are still debated, a number of consistent observations have been presented in the literature [12, 32, 35, 36]. PEG is an uncharged, hydrophilic polymer, which is non-immunogenic and characterized by a high steric stabilization. Figure 6 shows the zeta potential of the investigated QDs in PBS (pH = 7.2), along with the average internalization rate over the investigated cell lines.

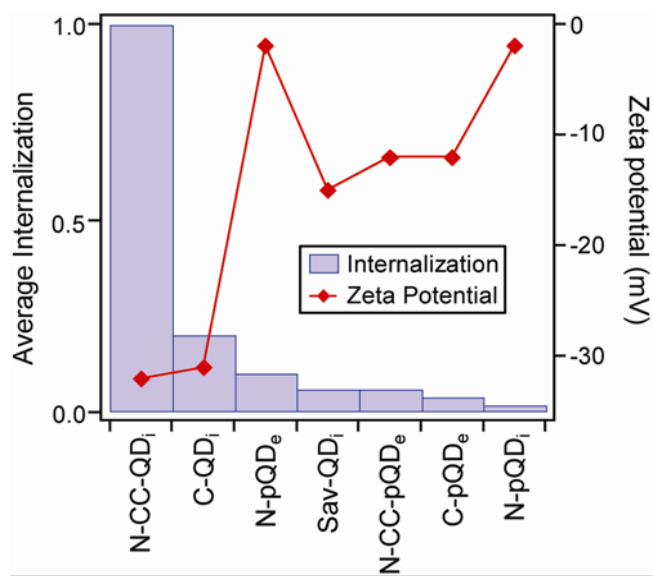

*Figure 6. Plot of the QD zeta potentials in PBS, pH 7.2, and average internalization levels.*

As can be seen, PEG-derivatized QDs had zeta potentials closer to zero than their pristine counterparts. This surface charge neutralization occurs due as the PEG molecules stopping the liquid flow in their vicinity. The fluid slippage plane was, therefore, shifted, and, as a result, modified the zeta potential (the zeta potential is defined the electrostatic potential measured at the slippage plane – the further the slippage plane from the particle surface, the smaller the potential, and hence, the zeta-potential ) [14, 18]. This effect is influenced by both PEG chain length, and the density of PEG molecules on the nanoparticle surface. To test this dependency, we performed EDC reactions to conjugate two different length PEG molecules (molecular weight 3400 and 5000 Da) in different surface concentrations to C-QD$_i$, as shown in figure 7.

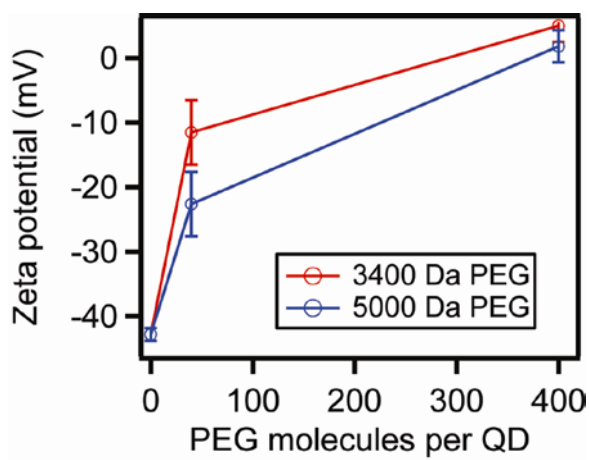

*Figure 7. Zeta-potential of QDs derivatized by 3400-Da and 5000-Da PEG molecules at different surface densities.*

We calculated that up to 400 PEG molecules can covalently bind to an individual QD. At this PEG surface density the measured zeta potential was close to zero, showing almost complete screening of the QD surface charge. A PEG layer thickness of 2 nm was inferred from the hydrodynamic diameter measured by DLS. As the surface density was reduced, the surface became more highly charged (the zeta potential shifted to the negative), approaching the value for the uncoated C-QD$_i$. The PEG surface density was correspondingly reduced, as confirmed by the DLS measurements. Note that 3400-Da and 5000-Da PEG-derivatization showed similar result within the experimental accuracy, in agreement with the literature reporting the effect of PEG-derivatization on the internalization at the 660-Da molecular weight [12].

A balance between the PEG length and surface density is important for optimal design of nanoparticle PEG-coating. When densely packed, PEG covers the majority of surface binding sites, and the molecules align perpendicular to the particles surface. At a lower density, due to the flexibility of PEG, it conforms to a 'mushroom' type state, where the chains are bent over the surface, again covering a large number of surface sites [18]. The main difference between the two conformational states is that at the lower density, the PEG chains tangle and cause the terminal groups to become covered with the PEG layer. As a result, non-specific internalisation [32, 41] is reduced. By the same token, a number of bioconjugation-active terminals are also reduced. While only limited information about the precise nature of the different QDs has been provided by the manufacturers, a different PEG length and density ratio is likely to account for the difference in internalization between N-pQD$_i$ and N-pQD$_e$.

It has been shown that charge has a profound effect on the rate of non-specific internalization [9, 40], through electrostatic interactions. The PEG molecules reduce the ability of the QD to undergo these interactions by physically separating the surface from that of the cell (steric hindrance), hence reducing QDs adherence to the cell membrane [13, 22, 41], and leading to reduced internalization. PEG also acts efficiently to reduce protein adsorption onto the nanoparticle surface [12]. This property has been shown prevent phagocytosis [11, 12], as well as reducing the binding of nanoparticles to the cell membrane, and hence lowering internalization [16].

We now discuss the surface charge effect. The importance of surface charge on the non-specific internalization rate has been discussed in the past [12, 18, 19], and it has been shown that both positively and negatively charged molecules can be efficiently internalized into cell, utilizing interactions with various charged proteins on the cell membrane [9, 10, 40]. It has also been shown that nearly neutral QDs with hydroxyl (-OH) surface functionalization, also exhibit greatly reduced non-specific internalization [13], when compared with the carboxyl or amino counterparts, so charge is of a definitive importance to understand the internalization characteristics.

As noted above, presence of the PEG-layer reduces an effective number of the terminal groups due to the entanglement of the chains, and hence it reduces the apparent charge. The zeta-potential measurements showed that PEG-derivatized QDs with amino terminal groups, while still negative, were charged more positively in comparison with those of the carboxyl groups, as expected. This negative charge is attributed to the intrinsic zeta-potential of the pristine QDs. This was confirmed by the measurements of the PEG-free QDs, which both showed large negative zeta-potentials regardless of the terminal groups.

Positively charged amino-functionalized particles exhibit the greatest internalization rates [10, 42], that can be ten-fold higher than negatively-charged carboxylated particles in neutral pH [40]. This is in agreement with our results, where both N-CC-$QD_i$ and C-$QD_i$ were internalized to a greater extent than SA-$QD_i$ and amino-functionalized particle internalization rate was, on average, ten-fold that of their carboxyl equivalents. It should be noted that the overall charge of the QD will influence electrostatic interactions with the cell membrane, although molecular interactions with individual proteins become important in close proximity. We propose that that this is the reason that N-CC-$QD_i$ internalization exceeds that of the C-$QD_i$, while the zeta potential is similar.

**Summary**

The level of cellular internalization of various commercially available QDs by three different tumour cell lines has been investigated. Overall, we showed that the presence of PEG on the particle surface significantly reduced the internalization. This was due to the PEG-layer-induced steric hindrance: the direct docking of QDs to the cellular membrane was obstructed, as well as protein adsorption on the QD surface, which otherwise would promote internalization. Charge has the greatest effect on internalization level, so should be most carefully considered when conducting culture experiments in live cells. While the uptake mechanisms of nanoparticles into cells is defined as non-specific, these pathways were initiated by surface receptors, either through direct interactions between the charged particle and the receptor or via proteins adsorbed on the nanoparticles surface.

Commercially available, PEG-derivatized QDs are appropriate as control particles due to their low non-specific binding/internalization into the investigated cell types, of these we find N-p$QD_i$ to be best, showing almost zero levels of internalization. Amino or carboxyl, PEG-free QDs can be absorbed by the cells in relatively large quantities due to their large surface charges, and while both positive and negatively charged particles are effectively internalized, positive amino capped particles are the most efficient. Such particles can serve as useful staining models, or for toxicity evaluation, where pathway-specific trafficking into the cell is of little importance. The importance of particle size should also be carefully considered, particles of the order of 50 nm will be most efficiently internalized into the cell, but this must be balanced with need of the particle once internalized, for protein pathway tracking for example, this size will affect the proteins movements in the cell.

This paper highlights that non-specific uptake is an important factor, when considering the use of nanoparticles in cellular systems. We believe that these observations will be of use to researchers moving from the fields of physics and nanotechnology into cell biology in providing the basis needed to perform efficient studies into nanoparticle internalization and interactions with cellular systems.

**Acknowledgment**
We appreciate help and equipment loans from David Inglis. Mark Connor, Medical Foundation of The University of Sydney, provided the AtT20 cells. The work was supported by the Macquarie University Research Innovation Fund #1136900.